\documentstyle[preprint,aps,epsf]{revtex}

\newcommand{\be}{\begin{equation}}
\newcommand{\ba}{\begin{eqnarray}}
\newcommand{\ee}{\end{equation}}
\newcommand{\ea}{\end{eqnarray}}

\def\pra#1#2#3{{\it Phys.~Rev.\/}~{\bf A#1} (19#2) #3}

\begin{document}

\preprint{\setlength{\baselineskip}{1.5em}\vbox{\hbox{}}}

\draft

\title{Coherent X-Ray Generation with Laser Driven Ions}

\author{Massimo Casu$^{}$, Carsten Szymanowski$^{a}$,
Suxing Hu$^{b}$ and Christoph H. Keitel$^{}$}

\address{Theoretische Quantendynamik, Fakult\"at Physik, Universit\"at Freiburg\\
Hermann-Herder-Str. 3, D-79104 Freiburg, Germany\\
 email: casu@physik.uni-freiburg.de or keitel@uni-freiburg.de \\
$^a$ New Address: Department of Physics and Astronomy, University of Rochester, Rochester NY 14627\\
$^b$ New Address: Max Born Institut, Rudower Chaussee 6, D-12489 Berlin, Germany}
\date{\today}

\maketitle

%%%%%%%%%%%%%%%%%%%%%%%%%%%%%%%%%%%%%%%%%%%%%%%%%%%%%%%%%%%%%%%%%%%%%

\begin{abstract}
Small parts of the electronic wavepacket of a  multiple charged ion 
may still tunnel through the high Coulomb barrier for a sufficiently 
intense laser field of $10^{15}W/cm^2$ and higher. Solving numerically the
corresponding Schr\"odinger equation we find that the periodic 
recollisions of such a wavepacket in the oscillating high power laser 
field will give rise to coherent X-ray radiation in the multiple keV regime,
i.e.\ substantially higher than predicted or observed before.    
\end{abstract}

%%%%%%%%%%%%%%%%%%%%%%%%%%%%%%%%%%%%%%%%%%%%%%%%%%%%%%%%%%%%%%%%%%%%%
\vspace{1.5cm}

%PACS Numbers:  42.65.Ky, 42.50.Hz, 32.80.Rm, 42.65.Re 
\newpage

\vspace{1.0cm}

%\section{Introduction}

Since the invention of the laser, scientists have tried to generate coherent
light also far below the optical wavelength \cite{Xreview}. The striking advantages 
in high resolution measurements and imaging are doubtless. All X-ray lasers up to 
date however still lack the efficiency, size, price and in particular coherence 
properties of optical laser schemes. In one of the most promising mechanisms 
for coherent high frequency generation, high power near optical laser fields were
imposed on atomic systems. Parts of the electron wavepackets are then oscillating
through the ionic core and due to the nonlinear interaction with the nucleus give 
rise to multiples of the applied laser frequency, i.e. harmonics \cite{Harmonics}. 
In the so called tunneling regime 
only very small parts of the wavepackets tunnel through the laser modified Coulomb
barrier but when returning the so far highest orders of harmonics were
observed \cite{Lhuillier,Macklin}. The highest possible achievable photon energy  
has been determined theoretically to give the so called cut-off law 
\cite{Krause,Becker,Corkum}
\be
\label{cutofflaw}
n_{max}\hbar\omega = I_p + 3.2 U_p
\ee
with the ionisation potential $I_p$, the ponderomotive energy
$U_p=E^2/4\omega^2$, $\hbar$ the Planck constant, $n_{max}$ the maximal order 
of harmonics and $E$ and $\omega$ 
being the maximal amplitude and angular frequency of the applied laser field.\\
The above simple formular was then explained in a simple three step model 
\cite{Lewenstein}: (i) parts from the bound electron wavepacket tunnel through the 
Coulomb barrier at the time when the laser field is close to its maximum and is
sufficiently narrow for a substantial tunnel rate (ii) the born wavepackets oscillates 
freely and classically in the laser field and when returning has picked up the kinetic 
or ponderomotive energy 3.2 $U_p$ and (iii) at the return with the nucleus may at best 
drop back to the ground state of energy $-I_p$ and release in form of radiation the 
ionisation energy $I_p$ and all the gained kinetic energy 3.2 $U_p$. If this process is 
repeated several times, interference among those processes leads to coherence of the 
generated light \cite{Review}.  

Many attempts have been made to enhance the above cut-off rule for the benefit of
even higher coherent high frequency generation. By preparing the atomic ensemble 
appropriately prior the interaction with the laser field or turning to more complex
systems as molecules, the factor 3.2 could be enhanced to at best the order of 
10 \cite{Review}. In principle the $U_p$ could be strongly increased by several orders 
of magnitude with present laser technology by increasing the driving laser
field strength up to the relativistic regime \cite{rel}.
However, the Coulomb barrier of neutral atoms is then modified so strongly that the 
above described three step model does not apply anymore and instead often only 
Bremsstrahlung harmonics arise.
  
A large experimental enhancement was recently achieved independently by the groups of Krausz 
\cite{Spielmann} and by that of Murnane and Kapteyn \cite{Chang} where ultra short pulses were 
employed. With at about 500 eV they have generated the shorted coherent wavelength so far
and have entered the so called water window where optical imaging of living species in water is 
possible. The break through was possible essentially because of the use of very short pulses
which allowed to move quickly to the highest intensity of the laser pulse before too much
of the ground state wavepacket was lost. In spite of this dramatic improvement a 
draw back of the short pulses is that only few recollisions of born wavepackets with the 
atomic core are possible and thus the coherence properties of the generated light are limited.
   
In this Letter we present a way to enhance high harmonic generation towards the hard X-ray regime, 
maintaining still high coherence properties. For this purpose 
we propose to use multiply charged ions \cite{Ditmire,gsi} rather than atoms because 
they possess a considerably larger ionization potential $I_p$ and according to the 
cut-off rule Eq.(\ref{cutofflaw}) promise a far larger maximal harmonic order. Since relativistic
corrections turned out negligible for the low ion charges of interest here, we solve 
numerically the corresponding Schr\"odinger equation.  For each effective ion charge we evaluate
the appropriate laser electric field strength to allow for tunneling and recollision harmonic 
generation. For a charge of $Z$ = 3 of the ionic core as sensed by the active electron, 
harmonics up to the order 2500 arise, i.e.\ with a photon energy of more than 4 keV. 

The parameter that has been employed so far to estimate if the laser and atomic
parameters are such that the three step tunnel mechanism takes place is the  
Keldysh parameter $\gamma_K$ \cite{Review}
\be
\label{keldysh}
\gamma_K = \sqrt{\frac{Z^2 I_p^{(1)}}{2 U_p}} = \frac{Z\omega}{E}\sqrt{2I_p^{(1)}}
= \frac{Z\omega}{E}
\ee
with $I_p= Z^2 I_p^{(1)}$ being the ionization potential of the ion of interest with 
effective charge Z and $I_p^{(1)}$ the ionization potential for hydrogen.
If this parameter is well below unity, one is in the tunneling regime. To have 
significant tunneling and thus harmonic intensity, we need a small $\gamma$. However
to avoid substantial ionization this parameter should not be too small.
Following ref. \cite{Amm} tunneling decreases exponentially with $Z^3$ and a $U_p$
increasing with $Z^6$ is necessary to compensate for this.
 As this is only an estimation for our model potential we have considered and optimized
the tunneling dynamics also via numerical studies prior calculating the spectrum.  
We show in fact that we have to decrease the $\gamma$ parameter with increasing
charge of the ion strongly to ensure a sufficiently large tunneling rate and thus high 
harmonic output. 

The laser field strength employed do not exceed the order of
$2 \cdot 10^{16}W/cm^2$ so that due the laser field we do not enter the 
relativistic regime.  Concerning the interaction of the electronic wavepacket with
the ionic core we employ a net charge of up to $Z=3$. This can also be assumed as 
non relativistic; a corresponding calculation via the Dirac equation will
thus give essentially equal results.
For more highly charged ions the velocities close to the nucleus can be significant
as compared to the speed of light $c$ so that the laser Lorentz force term 
can be nonnegligible as compared to the laser electric field \cite{Hus}.
This term would then give rise to complex modifications to the recollision dynamics 
and thus to the harmonic spectrum but is not of interest to the nonrelativistic
regime of interest here \cite{Hu}. 

With those assumptions in mind we consider the following dynamical equation for
the electronic wavefunction $\Psi(x,t)$ as a function of the spatial coordinate $x$
in the direction of the linearly polarized laser light and the time $t$:  
\be
i\partial_t \Psi(x,t) = H_{S} \Psi(x,t)
\ee
where $H_{S}$ stands for the Schr\"odinger Hamiltonian with:
\ba
H_S &=& \frac{1}{2}\left ( p-\frac{A(t)}{c} \right )^2 + V(x,Z)
\ea
where the vector potential of the laser pulse is defined via
$A(t)=E\cdot h(t) \cdot\sin(\omega t)$. The electric field  $E \cdot h(t)$ involves
a pulse shape $h(t)$ with turn-on phase and a phase with constant amplitude during which we 
evaluate the radiation spectrum.

We use the regularized nuclear potential, 
 \be
V(x,Z) = -\frac{k}{\sqrt{x^2+s}}
\ee
where $s=2$ is a smoothing parameter of the nucleus to compensate for the 
neglect of the second and third dimension \cite{Eberly}.
$k$ is a direct function of the net charge $Z$ of the ion and can be adjusted 
to recover the experimental values for the binding energies. 
  
\begin{figure}[t]
\begin{center}
\unitlength1cm
 \begin{minipage}[t]{16cm}
  \begin{center}
   \mbox{\epsfxsize 16cm \epsffile{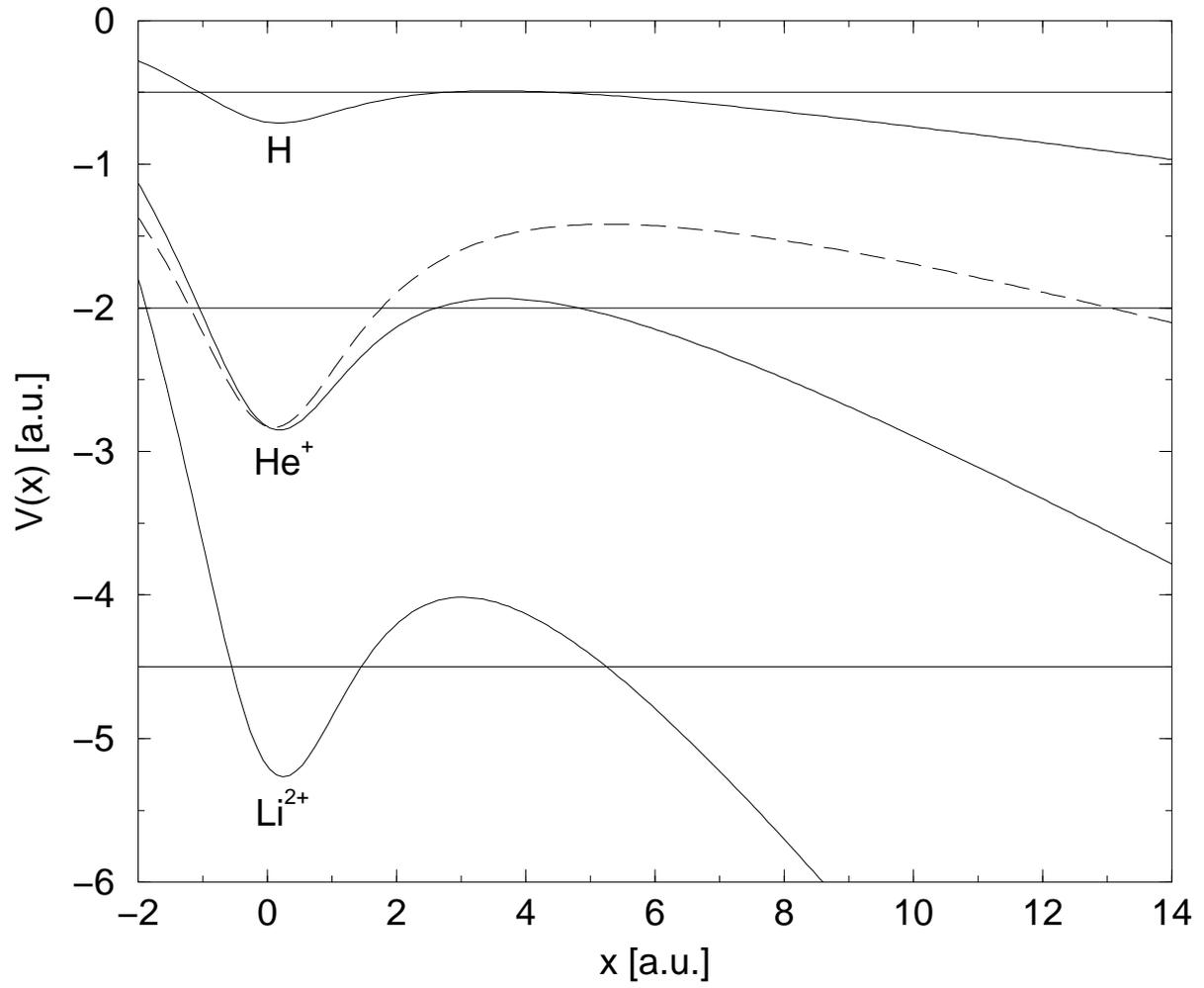}}
  \end{center}
 \end{minipage}
\end{center}
\caption{M. Casu et al., ``Coherent ...''}
\label{pot}
\end{figure}

In our numerical analysis for hydrogen (H) and hydrogen like ions (He$^+$, Li$^{2+}$) 
interacting with short intense laser pulses, we employed a laser wave length 
of $\lambda=800$nm (TiSapphire), an electric field  amplitude with intensity 
$I \sim 10^{14}-10^{16}$W/cm$^2$ and pulse lengths between 20 and 200 cycles with 
a turn-on phase of  5 to 20 cycles. 
In fig. \ref{pot} we displayed the effective potentials for H, He$^+$ and Li$^{2+}$ 
in the laser pulse at the time of maximal electric field and as a horizontal
line the corresponding ground state energy. 
For the case of hydrogen we have chosen the laser parameters ($\gamma_K=0.89$) such that
 we are well in the tunneling regime but simultaneously such that ionization is still small.
When moving to $He^+$ we first increased the electric field strength by a factor of two only 
to obtain the same Keldysh parameter and note from the corresponding dashed line for the
effective potential in Fig. 1 that the tunneling barrier is then very large and that we appear
to be  in the intermediate regime to the multiphoton regime. 
While the ionization potential scales with $Z^2$, the tunneling rate reduces expotenially with 
$Z^3/E$ \cite{Amm}.
In the solid line for $He^+$ we have further increased the electric field strength such that
we have similar ionization as for $H$ above, however though with a smaller $\gamma_K$.

Before addressing $Li^{2+}$, we discuss Fig. 2 where we have displayed the radiation spectrum
corresponding to the situations in H (a) and the two cases for $He^+$ displayed in Fig. 1. For H we 
are in the tunneling regime and thus find as well known a plateau and cut-off following Eq. (1). 
In Fig 2(b) for increasing $I_p$ and $U_p$ by a factor of two we find a spectrum with the 
expected cut-off enhanced by a factor of $Z^2=4$.  However, even though $\gamma_K$ is unchanged
we have clearly moved towards the multi photon regime. The harmonic structure has detoriated and
we find even resonance structures in the low frequency part of the spectrum. The enhancement of
about the 12th harmonic corresponds to the transition from the first excited to the ground state.
Also we note a beginning of a tilting of the spectrum in the plateau area.
To reenter the tunneling regime, we have increased the E field further, such that the ionization 
rate becomes comparable to that for $H$ in Fig. 1 a) and found an up-conversion of the 
cut-off frequency by a factor of 10.

We continue the procedure for $Li^{2+}$ as depicted in the lowest entry of Fig. 1. We have increased 
the electric field with respect to $H$ not only by a factor 3 but substantially more to secure 
sufficient tunneling. However, for such high harmonics as expected here and the corresponding 
low efficiency we chose a long pulse for a clear resolution of the very high harmonics. This 
requires us to maintain a sufficiently wide ionization barrier somewhat larger than in the two previous 
cases and thus accept a spectrum with a less pronounced cut-off frequency. Still we achieve at 
about 2500 harmonics as visible in Fig. 3, i.e. hard X-ray harmonics with an energy of at about 
4 keV. We note that sometimes in the literature those energies are still counted to the upper end 
of soft X-rays. However, proceeding to higher charged ions, even higher harmonics are achievable, 
though with an increasingly small efficiency.
Given the available parameters we note a scaling law of at about $Z^3$ for the energy of the cut-off 
harmonics as a function of the ion charge and the correspondingly chosen laser field intensity.
We note that for practical reasons we have increased the intensity with rising charge $Z$
rather than modifying the frequency which is less controllable experimentally. A constant wavelength
of the applied laser field may however be problematic, since relative to the modified characteristic 
length of the ionic core due to the variable $Z$, one may need adapt the time available for tunneling.

Moving towards the relativistic regime \cite{rel} with higher laser intensities and ionic charges, higher harmonics 
are possible. However, the harmonic spectrum deviates further from the traditional structure in the tunneling regime
including a horizontal plateau and a well pronounced cut-off. The magnetic component of the laser field induces a 
significant momentum transfer in the propagation direction of the laser field. This makes recollisions more 
difficult, especially involving long return times after many free oscillations in the laser field.   
As a consequence we find a reduction of harmonics between the perturbative and the cut-off regime. This goes beyond
the scope of this letter and will be discussed in detail in future work \cite{Hu}.

In conclusion multiply charged ions still allow for tunneling and recollisions of electron wave packets
with the parent ionic core with laser field intensities well above those employed before for high 
harmonic generation with atoms. Consequently coherent X-rays become feasible in the multiple keV 
regime. The efficiency is below that for coherent high harmonic generation via laser driven atoms.

\vspace{0.5cm} 

This work has been funded by the German Science Foundation (Nachwuchsgruppe within 
SFB 276). CS and SXH acknowledge present funding from the Alexander von Humboldt foundation.

%%%%%%%%%%%%%%%%%%%%%%%%%%%%%%%%%%%%%%%%%%%%%%%%%%%%%%%%%%%%%%%%%%%%%

%%%%%%%%%%%%%%%%%%%%%%%%%%%%%%%%%%%%%%%%%%%%%%%%%%%%%%%%%%%%%%%%%%%%%

\newpage 

\begin{center}
{\large Figure Captions}
\end{center}

\begin{itemize}
\item[Fig.1.] The effective potentials for $Z=1$ (H, k=1),
$Z=2$ (He$^+$, k=3.48) and $Z=3$ (Li$^{2+}$, k=7.35) interacting with 
 electric fields corresponding to the intensities 
$1.4\cdot 10^{14}$W/cm$^2$ ($\to \gamma_K=0.89$ for H), 
$2.19\cdot 10^{15}$W/cm$^2$ ($\to \gamma_K=0.456$ for He$^+$) and 
$2\cdot10^{16}$W/cm$^2$ ($\to \gamma_K=0.255$ for Li$^{2+}$).
The corresponding ground state energies are indicated by the horizontal lines. 
The laser wavelength is $800$nm, which corresponds to an atomic frequency 
$\omega=$0.057au. He$^+$ with the same $\gamma_K$ as $H$ is represented by the dashed
line.
\item[Fig.2.] 
The harmonic spectrum for H (a) and He$^+$ (b/c) in appropriately intense laser fields
as given in Fig. 1.  
The pulses consist of  20 cycles full intensity and 5 cycles turn-on. With
Eq. \ref{cutofflaw} the cut-off is placed at $n_{max,H}=27$ for H, $n_{max,He^+}=Z^2 n_{max,H}=108$ 
for He$^+$ with the same $\gamma_K$ as for H and $n_{max,He^+}=305$ with the enhanced 
$U_p$ (the ionisation rate here is comparable to that of H and is about 20\%).
\item[Fig.3.]
a) The harmonic spectrum for Li$^{2+}$ for the laser parameters given in Fig. 1. 
b) is an enlargement of the high frequenct range of figure a).
The full intensity was applied here for 150 cycles with an 18 cycle turn-on. After 100 cycles
we have $\sim 60\%$ ionised and just after 150 cycles essentially 100$\%$ is ionised. 
\end{itemize}

%%%%%%%%%%%%%%%%%%%%%%%%%%%%%%%%%%%%%%%%%%%%%%%%%%%%%%%%%%%%%%%%%%%%%

	\begin{figure}[t]
\begin{center}
\unitlength1cm
 \begin{minipage}[t]{16cm}
  \begin{center}
   \mbox{\epsfxsize 16cm \epsffile{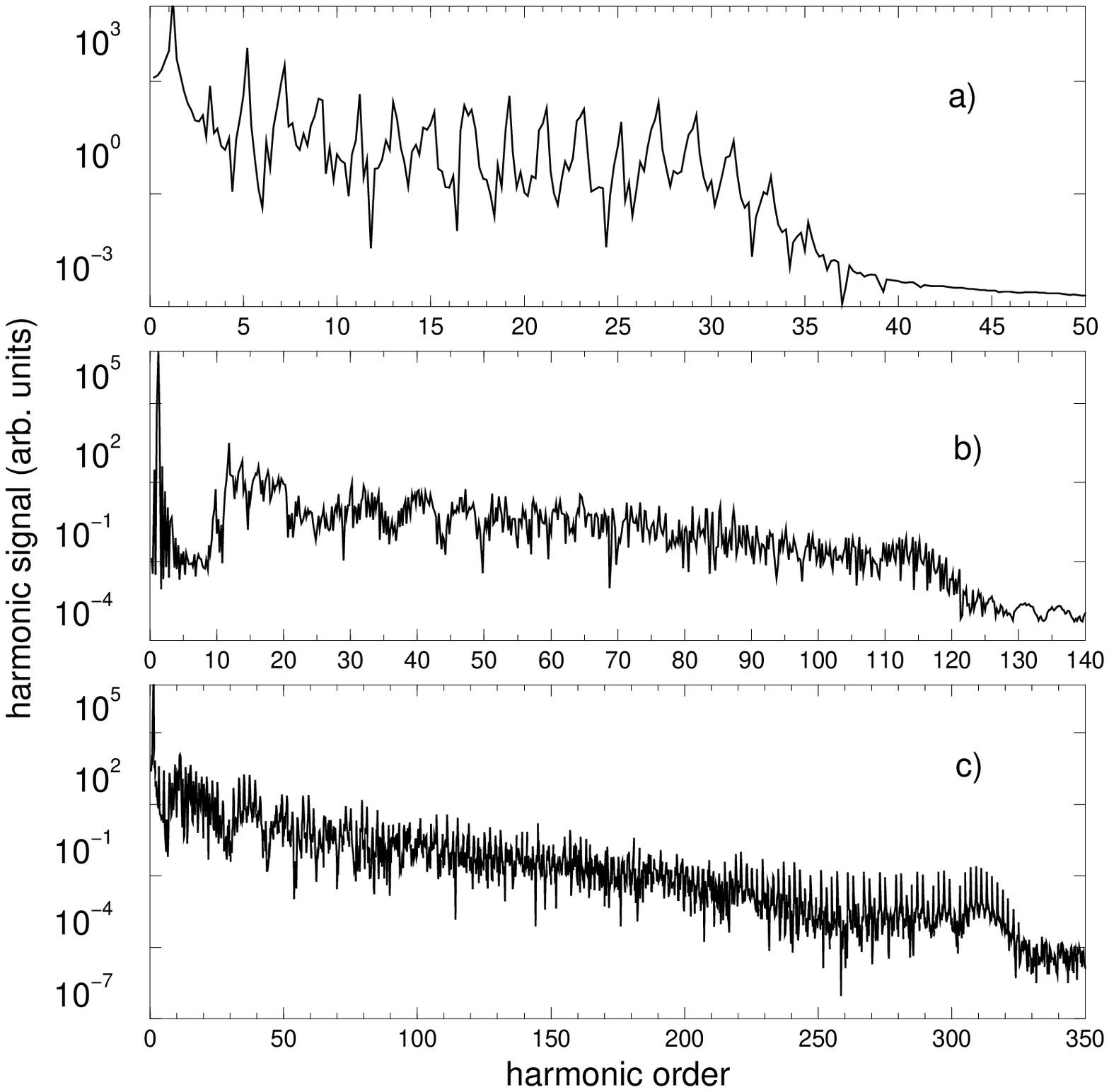}}
  \end{center}
 \end{minipage}
\end{center}
\caption{M. Casu et al., ``Coherent ..''}
\label{spec}
\end{figure}

\begin{figure}[t]
\begin{center}
\unitlength1cm
 \begin{minipage}[t]{16cm}
  \begin{center}
   \mbox{\epsfxsize 16cm \epsffile{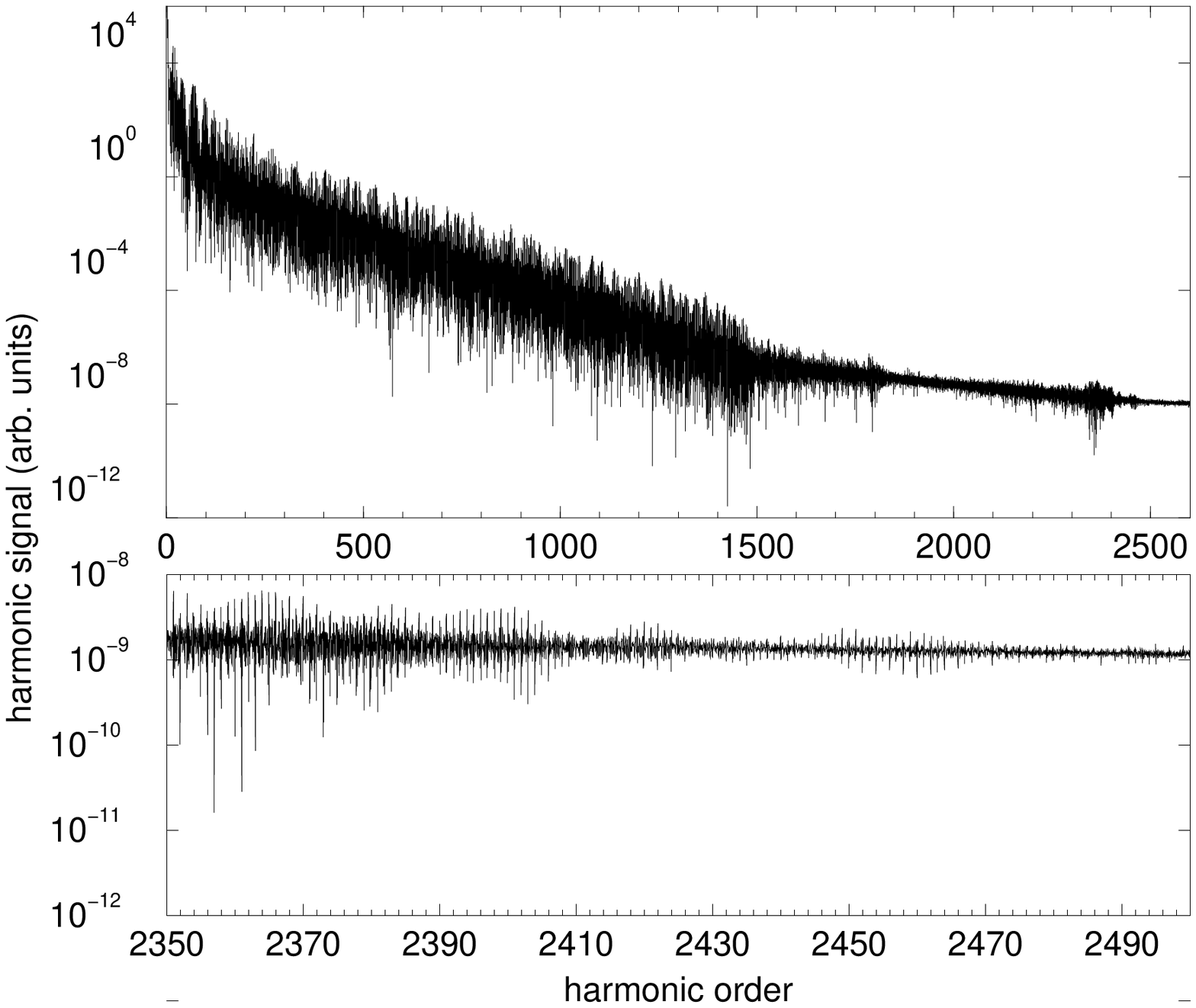}}
  \end{center}
 \end{minipage}
\end{center}
\caption{M. Casu et al., ``Coherent ..''}
\label{z3}
\end{figure}

\end{document}